\documentclass[10pt]{iopart}
\usepackage{graphicx}
\usepackage[normalem]{ulem}

\usepackage{color}

\begin{document}

\title[E Mazaleyrat \textit{et al.}]{How to induce superconductivity in epitaxial graphene \textit{via} remote proximity effect through an intercalated gold layer}

\author{Estelle Mazaleyrat$^{1,2,*}$, Sergio Vlaic$^3$, Alexandre Artaud$^{1,2}$, Laurence Magaud$^2$, Thomas Vincent$^3$, Ana Cristina G\'omez-Herrero$^2$, Simone Lisi$^2$, Priyank Singh$^2$, Nedjma Bendiab$^2$, Val\'erie Guisset$^2$, Philippe David$^2$, St\'ephane Pons$^3$, Dimitri Roditchev$^3$, Claude Chapelier$^1$, and Johann Coraux$^{2,\circ}$}
\address{$^1$ Univ. Grenoble Alpes, CEA, IRIG, PHELIQS, 38000 Grenoble, France}
\address{$^2$ Univ. Grenoble Alpes, CNRS, Grenoble INP, Institut N\'eel, 38000 Grenoble, France}
\address{$^3$ Laboratoire de Physique et d'\'{E}tude des Mat\'eriaux, ESPCI Paris, PSL Research University, CNRS, Sorbonne Universit\'es, UPMC Univ Paris 06, 75005 Paris, France}
\address{$^*$ ORCiD: 0000-0002-7447-7763}
\address{$^\circ$ ORCiD: 0000-0003-2373-3453}

\begin{abstract}
Graphene holds promises for exploring exotic superconductivity with Dirac-like fermions. Making graphene a superconductor at large scales is however a long-lasting challenge. A possible solution relies on epitaxially-grown graphene, using a superconducting substrate. Such substrates are scarce, and usually destroy the Dirac character of the electronic band structure. Using electron diffraction (reflection high-energy, and low-energy), scanning tunneling microscopy and spectroscopy, atomic force microscopy, angle-resolved photoemission spectroscopy, Raman spectroscopy, and density functional theory calculations, we introduce a strategy to induce superconductivity in epitaxial graphene \textit{via} a remote proximity effect, from the rhenium substrate through an intercalated gold layer. Weak graphene-Au interaction, contrasting with the strong undesired graphene-Re interaction, is demonstrated by a reduced graphene corrugation, an increased distance between graphene and the underlying metal, a linear electronic dispersion and a characteristic vibrational signature, both latter features revealing also a slight $p$ doping of graphene. We also reveal that the main shortcoming of the intercalation approach to proximity superconductivity is the creation of a high density of point defects in graphene (10$^{14}$~cm$^{-2}$). Finally, we demonstrate remote proximity superconductivity in graphene/Au/Re(0001), at low temperature.
\end{abstract}

\color{black}
%
\vspace{2pc}

\submitto{\TDM}
%
%

\ioptwocol

\section*{Introduction}
Except in the rare cases of magic-angle-twisted bilayer graphene heterostructures \cite{cao2018}, graphene is not intrinsically superconducting. Still, it can be rendered superconducting using the superconducting proximity effect, that has essentially been implemented ``from on top" of graphene, locally, by contacting superconducting electrodes \cite{heersche2007,du2008}, or \textit{via} spontaneously formed metallic superconducting micro- or nano-clusters \cite{feigel'man2008,kessler2010,allain2012,natterer2016}.

An alternative approach is to induce superconductivity in graphene ``from below", using a superconducting substrate, Re(0001) \cite{tonnoir2013}. The process by which graphene is made superconducting is here a truly bottom-up one, since no transfer step or electrode contacting is necessary. Graphene on metals is obviously not suited for electronic transport investigations and devices. However, it has offered insightful platforms to explore much sought electronic phenomena, superpotential effects \cite{pletikosic2009}, bandgap opening strategies \cite{balog2010}, straintronics effects \cite{levy2010}, and spin-orbit proximity effects \cite{varykhalov2008,calleja2015}, to name only a few. Unfortunately, unlike in these examples, the interaction between graphene $\pi$ and Re(0001) $5d$ states is strong \cite{tonnoir2013,gao2017}, hence there is no Dirac-like electronic band structure in this system \cite{papagno2013}. This has been a major limitation to the interest of the superconducting graphene/Re(0001) system so far.

Intercalation of a buffer atomic layer of certain elements in between graphene and its substrate is an efficient way to recover the intrinsic structural and electronic properties expected for isolated graphene. Obviously, the intercalated species should passivate the metallic substrate. In particular, graphene shows structural and electronic properties that are strongly reminiscent of isolated graphene (hence it is coined ``quasi free-standing") when the metallic intercalant has full higher-energy $d$ sub-shell, as shown with graphene/Ag/Re(0001) \cite{papagno2013}, graphene/Pb/Re(0001) \cite{estyunin2017}, graphene/Au/Ni(111) \cite{marchenko2012,shikin2013}, graphene/Cu/ Ni(111) \cite{shikin2013}, graphene/Ag/Ni(111) \cite{shikin2013}, graphene/Al/ Ni(111) \cite{voloshina2011}, graphene/Au/SiC(0001) \cite{premlal2009,gierz2010} and graphene/Pb/Ru(0001) \cite{fei2015}. 

Intercalation is however not neutral, and may have desired or spurious effects on graphene, including the possibility to adjust the electronic doping of graphene ``from below" \cite{calleja2015,leicht2014_acsnano,vita2014,varykhalov2010,marchenko2012,enderlein2010,klimovskikh2017,papagno2013,estyunin2017} and the possibility to generate defects \cite{huang2011,sicot2012,sicot2014}.

The choice of the intercalant can also be motivated by the wish to induce spin-orbit coupling in graphene: electronic bands with spin-splitting due to the spin-orbit proximity effect have been reported with Au \cite{varykhalov2008} and Pb \cite{klimovskikh2017,otrokov2018} intercalants. Strong spatial variations of proximity spin-orbit effects at the edge of intercalated islands were proposed to generate sharp Landau levels in graphene \cite{calleja2015}. The origin of these effects is not fully clarified, and may relate to the structure of the intercalated layer \cite{krivenkov2017,slawinska2019}.

Using epitaxial graphene strongly bound to its rhenium substrate as a starting point, we produce the first superconducting epitaxial graphene sheet with a preserved two-dimensional character by intercalating sub-monolayers to few layers of gold atoms in between graphene and rhenium. The quasi-free standing character of graphene is extensively investigated by means of electron diffraction (reflection high-energy, RHEED, and low-energy, LEED), scanning tunneling microscopy and spectroscopy (STM/STS), atomic force microscopy (AFM), angle-resolved photoemission spectroscopy (ARPES), Raman spectroscopy, and density functional theory (DFT) calculations. Beside establishing the quasi-free standing character of graphene and its unaltered rhenium-induced superconductivity, we show that graphene becomes slightly hole-doped (with the Fermi level 65~meV below the Dirac point) and that intercalation induces a significant density of point defects in graphene, amounting to $10^{14}$~cm$^{-2}$. Remote proximity superconductivity in intercalated epitaxial graphene opens a route for the exploration of exotic superconducting states, in the presence of spin impurities or spin-orbit effects.

\color{black}
\section*{Materials and methods}
Sample preparation was performed under ultrahigh vacuum (UHV) conditions (base pressure of 10$^{-10}$~mbar). Two separate UHV systems were used for sample preparation: one was equipped with a room-temperature UHV-STM and the other was used for ARPES experiments. 

A rhenium single-crystal cut with a (0001) surface and a thin Re(0001) film on sapphire were used as substrates. To grow graphene, we used: (i) repeated cycles of annealing/cooling down under C$_2$H$_4$ in the case of the Re(0001) single-crystal \cite{miniussi2014}, and (ii) high temperature exposure to C$_2$H$_4$ followed by slow cooling down in the case of the Re thin film \cite{tonnoir2013} (more details given in the supplementary information (SI)). These two procedures yielded, respectively: (i) $\sim$ 70\% surface coverage with graphene and a small fraction ($\sim$ 15\%) covered with rhenium carbide \cite{miniussi2014,mazaleyrat-to-be-published}, and (ii) $\sim$ 90\% surface coverage with graphene.

\begin{figure*}[!hbt]
	\centering
	\includegraphics[width=14.6cm]{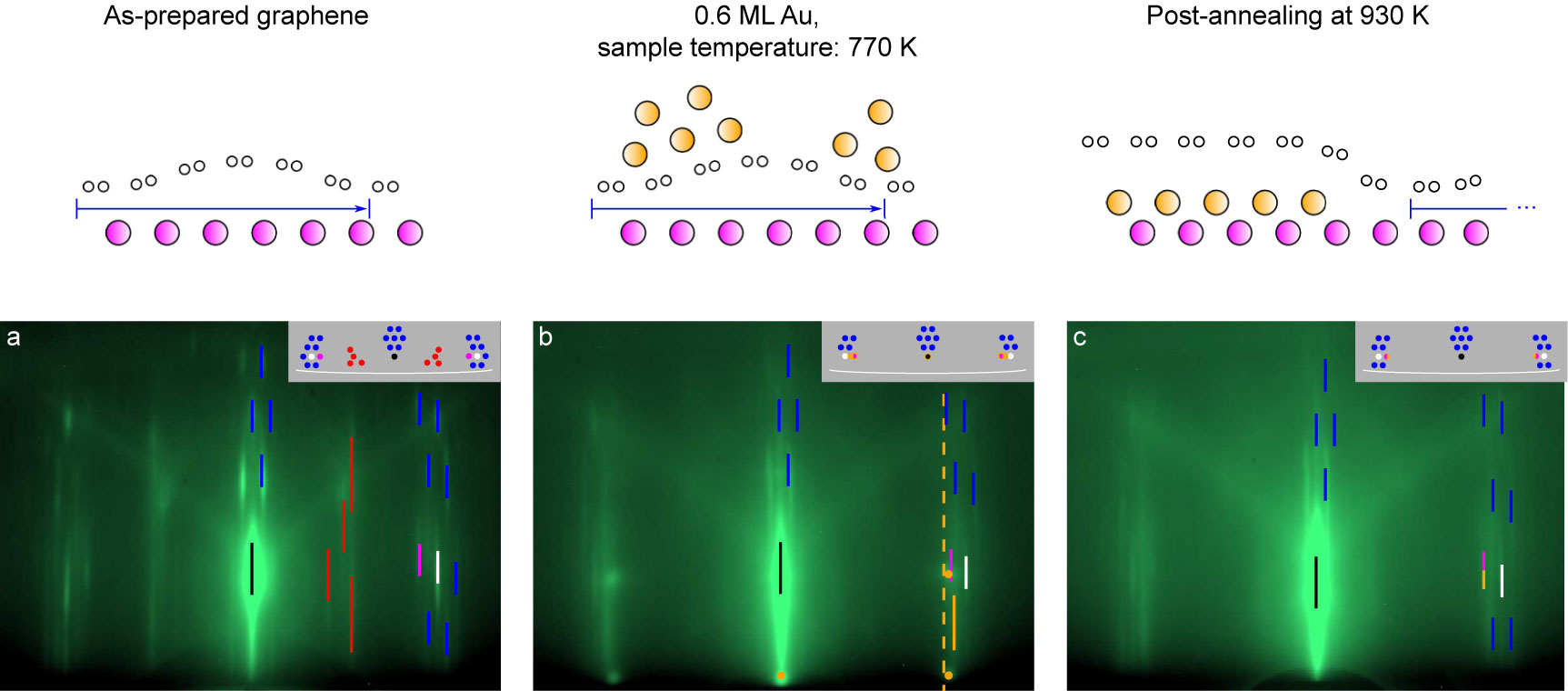}
	\caption{Top: Ball model of graphene-covered Re(0001) along the deposition/intercalation process of Au. Graphene, rhenium and gold atoms are schematically represented as white, pink and yellow balls, respectively. The moir\'e superperiodicity is indicated with blue arrows. Bottom: Periodic features observed with RHEED on (a) a graphene-covered Re(0001) surface, (b) after depositing 0.6~ML of Au at 770~K and (c) after subsequent annealing at 930~K. RHEED patterns were taken along the [01$\bar{1}$0] incident azimuth. Coloured lines appearing on the right-hand side of RHEED patterns indicate the specular (black), Re (pink), graphene (white), moir\'e (blue), rhenium carbide (red) and Au (yellow) streaks. The bulk lattice parameter of Au(111) is indicated with a dashed yellow line. Insets show schematic top views of the reciprocal space with all observed rods and a cut of the Ewald's sphere (white line).}
	\label{fig1}
\end{figure*}

The deposition rate of Au was estimated using either a quartz microbalance or Auger spectroscopy. Gold deposition and intercalation were performed as follows. First, gold deposition on graphene-covered Re(0001) was performed using either a Au evaporator or \textit{via} pulsed-laser deposition, at room temperature. Then, a subsequent annealing was performed at 970~K and 670~K in the two different UHV systems. An alternative method was used and gave similar results: Au deposition was performed with the sample held at 770~K, and a subsequent annealing at 930~K led to Au intercalation. 

ARPES experiments were performed \textit{in situ} using $\mathrm{He}_\mathrm{I}$ and $\mathrm{He}_\mathrm{II}$ discharge spectral lines (21.22~eV and 40.81~eV) as photon sources, at a sample temperature of about 100~K, using a SPECS PHOIBOS 225-2D-CCD hemispherical analyser (see more details in the SI). 

STM measurements were performed in two separate systems. In one of the same UHV systems where graphene was grown and intercalated with gold, STM was performed at room temperature using a commercial Omicron UHV-STM 1. The samples were also taken out of UHV, in a home-made STM/STS setup implemented inside an inverted dilution refrigerator working with a $^{3}$He/$^{4}$He mixture down to 50 mK. The differential conductance spectra were acquired using a standard lock-in technique (see more details in the SI). 

We used VASP \cite{kresse1993} to perform the \textit{ab initio} calculations, with the PAW approach \cite{kresse1999}, PBE functional \cite{perdew1996}, and Grimme corrections to van der Waals interactions \cite{grimme2006}. Four Re layers, one pseudomorphic Au layer, and a graphene layer were considered. The in-plane dimensions of the supercell correspond to a coincidence structure with (8$\times$8) graphene unit cells on top of (7$\times$7) Au/Re(0001). More details on the model used for DFT calculations are given in the SI.

Raman spectroscopy measurements were acquired with a 532 nm Nd:YAG laser using a commercial confocal WITEC spectrometer at room temperature under ambient conditions.

\section*{Gold intercalation process}
To optimize the Au intercalation process, we monitored the deposition of Au onto graphene/Re(0001) at 770~K, and the subsequent annealing of the sample to 930~K, with the help of real-time \textit{in situ} RHEED measurements (figure~\ref{fig1}).

Before Au deposition, the RHEED patterns consist of the characteristic streaks produced by diffraction by the graphene, rhenium and moir\'{e} lattices. The latter (super-)lattice corresponds to the 2.1~nm-periodicity coincidence lattice that stems from the lattice mismatch between graphene and Re(0001) (see also STM image in figure~S1). These streaks are highlighted with white, pink and blue lines in figure~\ref{fig1}(a). As expected in the case of a Re(0001) single-crystal used as substrate \cite{miniussi2014,mazaleyrat-to-be-published}, we also detect the diffraction streaks of the surface rhenium carbide (highlighted with red lines in figure~\ref{fig1}(a)), which is a minority carbon phase.

\begin{figure*}[!hbt]
	\centering
	\includegraphics[width=17.2cm]{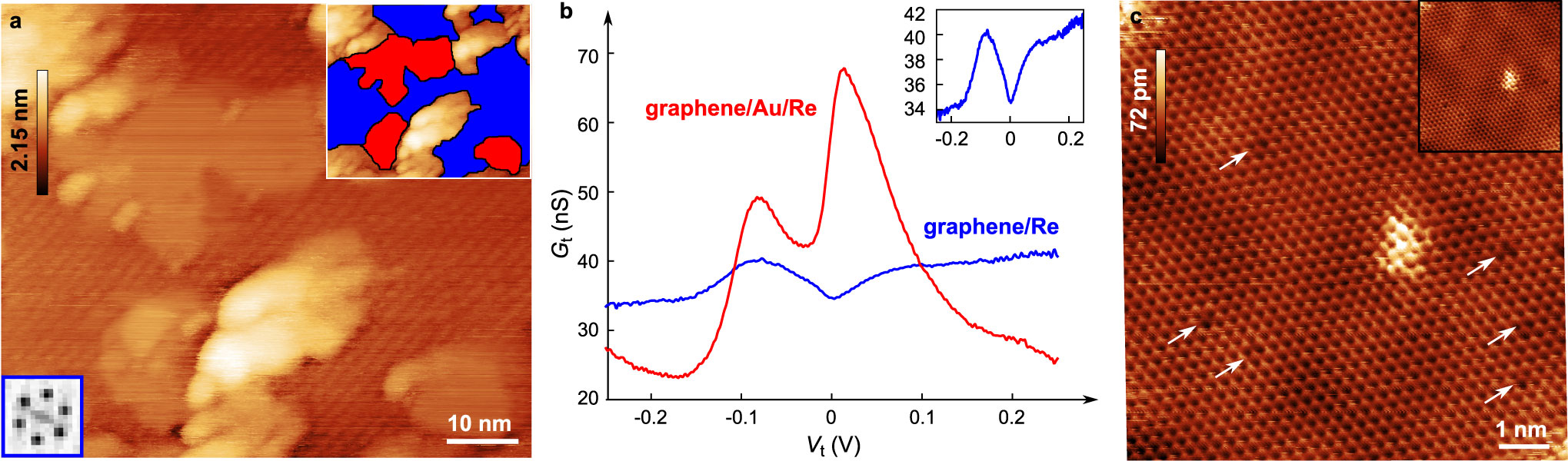}
	\caption{(a) Scanning tunneling topograph of graphene on Re(0001) after the deposition and intercalation of 0.6~ML of Au, revealing three types of regions: the moir\'e pattern characteristic of graphene on Re(0001), gold-intercalated graphene on Re(0001), and regions with gold clusters on top of graphene. $I_\mathrm{t}$ = 5~nA, $V_\mathrm{t}$ = -250~mV, T = 100~mK. Insets: (Top right) Scanning tunneling topograph overlaid with blue and red masks corresponding to graphene/Re and graphene/Au/Re regions, respectively. (Bottom left) (12~nm$^{-1}$~$\times$~12~nm$^{-1}$) Fast Fourier transform of a moir\'e region elsewhere on the surface, revealing the first moir\'e harmonics. (b) Conductance $G_\mathrm{t}(V_\mathrm{t})$ spectra measured for graphene/Re and graphene/Au/Re. T = 100~mK. Inset: View with a different $y$-scale of the conductance $G_\mathrm{t}(V_\mathrm{t})$ spectrum of graphene/Re. (c) Scanning tunneling topograph with atomic resolution of graphene on Re(0001) intercalated with 0.6~ML of Au. Defects in the graphene lattice are highlighted with white arrows. $I_\mathrm{t}$ = 10~nA, $V_\mathrm{t}$ = -1~V, T = 100~mK. Inset: Same scanning tunneling topograph. The moir\'e pattern is better visualized.}
	\label{fig2}
\end{figure*}

After depositing 0.6 monolayer (ML) of Au (with 1~ML denoting a 100\% coverage with Au pseudomorphic to Re(0001)) at 770~K (figure~\ref{fig1}(b)), some of the moir\'e streaks have disappeared, as well as the carbide streaks. This is an indication that gold has been deposited onto the surface. Besides, diffraction spots and streaks (yellow) appear after Au deposition (see SI movie~S1 and figure~\ref{fig1}(b)). The spotty diffraction signals correspond to Au deposited in the form of clusters, while the streaks rather point to the presence of a two-dimensional layer of gold (see more details in the SI).

The subsequent annealing at 930~K (figure~\ref{fig1}(c)) leads to the disappearance of gold spots. In the meantime, the Au-related streaks narrow, and their positions shift to match that of the Re streaks (see SI movie~S2). In other words, Au is now pseudomorphic to Re(0001) and structurally less disordered than prior to annealing. A straightforward interpretation is that Au is intercalated between graphene and Re(0001), and that the latter imposes its lattice parameter to the Au layer \footnote{As discussed later in the text, the free-standing character of graphene on Au/Re(0001) is synonymous of a very weak moir\'e pattern, with almost negligible electronic and structural spatial modulations. The fact that we do not observe a total disappearance of these streaks is presumably due to a smaller Au surface coverage (60\%) compared to the graphene coverage (70\%), which leaves pristine graphene/Re(0001) (with a moir\'e pattern) regions even after full intercalation of the deposited Au.}. This pseudomorphism is what is also observed in most metal-intercalated systems studied so far \cite{sicot2012,rougemaille2019}.

\section*{Structural and electronic properties of gold-intercalated graphene on Re(0001)}
\subsection*{Scanning tunneling microscopy and spectroscopy}
Figure~\ref{fig2}(a) shows a typical STM image of graphene/Re (0001) partially intercalated with Au. It exhibits three types of regions: (i) graphene/Re(0001) regions featuring the characteristic moir\'e superlattice (also detected in RHEED and LEED, see figures~\ref{fig1}(a) and S1(b)) with a superperiodicity in the nanometer range (in figure~\ref{fig2}(a) the moir\'e supperlattice appears as a pattern of lines due to the convolution with the shape of the tip; the inset of figure~\ref{fig2}(a) reveals a clearer moir\'{e} signature in the Fourier transform of another STM image) (ii) gold-intercalated graphene/Re(0001) regions forming 20-40 nm-large areas which appear much flatter (no visible moir\'e pattern in figure~\ref{fig2}(a)) than graphene/Re(0001), with a characteristic contrast in the phase signal in atomic force microscopy (see SI); (iii) gold clusters on top of graphene having a high aspect ratio (typically 1-1.5~nm and 15~nm in height and width) and presumably imaged multiple times by the tip for this reason (\textit{ex situ} AFM imaging provides a more reliable estimate of the surface density of these clusters, of 10$^{10}$~cm$^{-2}$, see SI). We disregard the latter regions in the following, and focus on the two other types of regions, namely graphene/Re(0001) and graphene/Au/Re(0001), which are highlighted in blue and red, respectively, in the inset of figure~\ref{fig2}(a). Gold-intercalated regions are 0.2-0.4~nm higher than non-intercalated regions, which corresponds to 1-2 intercalated gold monolayers. 

Figure~\ref{fig2}(c) shows a STM topograph with atomic resolution of an intercalated region. A shallow moir\'e superlattice with a superlattice parameter of $\simeq$~2.3~nm and a corrugation of approximately 80~pm (\textit{i.e.} about two times less than in the graphene/Re(0001) moir\'e) is visible. The observed periodicity is very close from that of the graphene/Re(0001) moir\'e, confirming that intercalated Au is pseudomorphic to Re(0001). The much weaker graphene corrugation is a signature of the much smaller graphene-Au interaction, compared to the graphene-Re interaction. 

The local conductance was measured at different locations on the STM image shown in figure~\ref{fig2}(a). A marked conductance peak is observed in graphene-covered Re(0001) regions around -0.1~eV (figure~\ref{fig2}(b)). In gold-intercalated regions, a resonance is also found at -0.1~eV. This resonance is stronger and it is accompanied by a second one, centered closer to the Fermi level (figure~\ref{fig2}(b)). The -0.1~eV resonance was previously observed in graphene/Re(0001) at the location of moir\'e hills and was attributed to an enhanced density of states in rhenium and carbon at the location of moir\'e hills, as evidenced by DFT calculations (Ref.~\cite{artaud2020}, see below). The enhanced density of states at the location of moir\'e hills results in an enhanced conductance in spectroscopy measurements. At the location of moir\'e valleys, such an enhanced conductance is not observed, probably due to the hybridization between the graphene and the metal states.

In the data shown in figure~\ref{fig2}(a,b), we do not resolve the moir\'e-site dependent variations of the local density of states reported in Refs.~\cite{tonnoir2013, artaud2020} (spectroscopy measurements were acquired at 1.5~nm distances from one another and figure~\ref{fig2}(b) shows the averaging of those spectra for graphene/Re and graphene/Au/Re regions). The persistence of the -0.1~eV resonance in gold-intercalated regions suggests that graphene at the location of moir\'e hills and gold-intercalated graphene are somehow similar, \textit{i.e.} they both exhibit a quasi free-standing character. 

\subsection*{DFT calculations}
To discuss the possible origin of the extrema observed in the scanning tunneling spectroscopy data we make first-principles calculations of the local density of states. Match between the experimental and calculated energy positions is expected within no better than a few 100~meV given the deviation from the (necessarily) simple and ideal model that we use for the simulations (see \textit{Materials and Methods}).

Our analysis of graphene/Au/Re(0001) should be compared to that made for graphene/Re(0001), which was reported in Ref.~\cite{artaud2020}. In the former system, graphene is much flatter than in the latter system: graphene's corrugation (taken as the maximum value of vertical separation between C atoms) amounts to 33~pm (\textit{versus} 170~pm for the latter system), consistent with the STM measurements (80~pm). The DFT calculations predict an average distance between the graphene layer and the Au layer of 3.24~\AA. This value is well beyond the distance of covalent carbon-metal bonds, indicative of weak bonding between graphene and the metal surface. The distance is similar to the value locally expected at the moir\'e hills in graphene/Re(0001), where an absence of C-metal hybridization is expected \cite{gao2017}.

\begin{figure*}[!hbt]
	\centering
	\includegraphics[width=17.0cm]{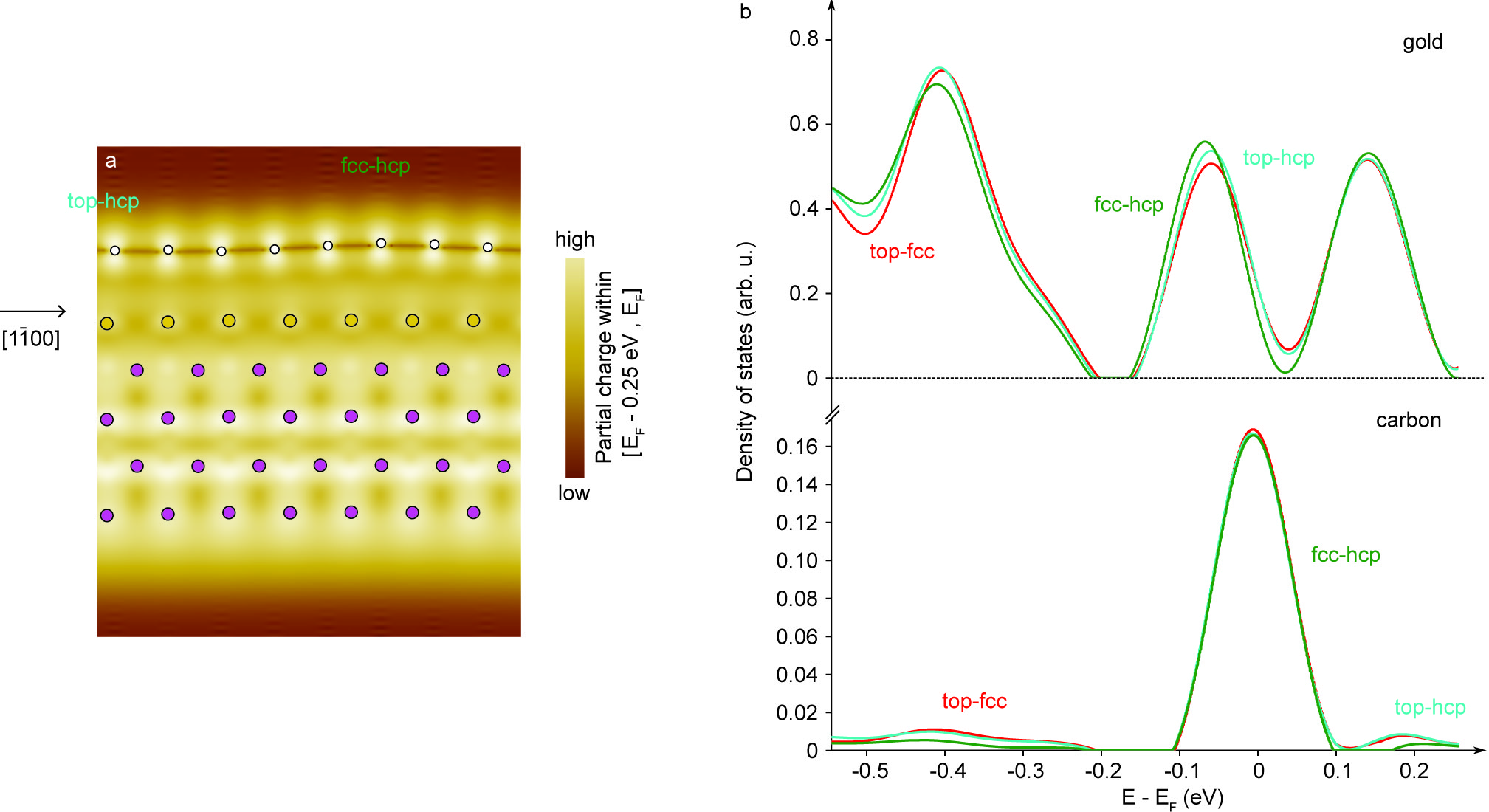}
	\caption{(a) Cross-section of the partial charge integrated between the Fermi level ($E_\mathrm{F}$) and $E_\mathrm{F}-0.25$~eV, along the [1$\bar{1}$00] direction, deduced from the DFT calculations on graphene/Au/Re(0001). Graphene, rhenium and gold atoms are schematically represented as white, pink and yellow balls, respectively. (b) Electronic density of states (electron energy $E$) averaged over a few Au atoms (one and three respectively, top curves) and averaged over a few C atoms (four and six respectively, bottom curves) at the hills ($fcc-hcp$) and valleys ($top-fcc$, $top-hcp$) of the moir\'e, from DFT calculations.}
	\label{fig3}
\end{figure*}

Contrary to graphene/Re(0001) \cite{artaud2020}, the extension of the Re electronic orbitals in the out-of-plane direction is essentially independent of the position in the moir\'e unit cell (figure~\ref{fig3}(a)). This extension is globally larger than on graphene/Re(0001), and comparable to the maximum extension in the latter system, that occurs at the moir\'e hills. This finding further indicates a similar quasi free-standing character of graphene in graphene/Au/Re(0001) and at the moir\'e hills of graphene/Re(0001).

Figure~\ref{fig3}(b) shows the density of states deduced from DFT calculations on graphene/Au/Re(0001). Similarly to the behaviour observed for the partial charge, the spatial variation of the density of states on Au atoms (figure~\ref{fig3}(b), top curves) and C atoms (figure~\ref{fig3}(b), bottom curves) is negligible. In particular, we find no significant variation in the energy positions of the observed maxima with respect to the local atomic stacking. 

Regarding the density of states on C atoms, one marked peak around the Fermi level is found in the calculations (with no additional feature). In agreement with the conclusion drawn in Ref.~\cite{artaud2020} for graphene-covered Re(0001), we suggest that this peak is at the origin of the enhanced conductance observed at -0.1~eV in gold-intercalated graphene on Re(0001) (see SI for further details on the analysis of the conductance spectra with/without intercalated Au).  

Regarding the density of states on Au atoms, two marked maxima at about -0.08~eV and 0.15~eV are observed. We suggest that the higher-energy maximum is responsible for the intense close-to-Fermi resonance, which is only present in gold-intercalated graphene, as shown in figure~\ref{fig2}(b). The lower-energy maximum is absent from conductance measurements. To rationalize the observation of only one of the two maxima of the Au density of states in the experimental conductance, we invoke an ``electronic transparency" of graphene only in the energy range corresponding to the highest-energy Au density of states maximum. This is reminiscent of the transparency reported in other epitaxial graphene systems \cite{riedl2007,brar2007,rutter2007,mallet2007,hiebel2009,schumacher2014}. This is also supported by our own DFT calculations, which predict a vanishing electronic density of states in the overlying graphene layer (figures~\ref{fig2}(b) and S3).

Overall, the two resonances observed in conductance measurements performed on gold-intercalated graphene on Re(0001) at -0.1~eV and close to the Fermi level are assigned to maxima appearing close to the Fermi level in the C density of states, and around 0.15~eV in the Au density of states, respectively.

\section*{Quasi free-standing character of gold-intercalated graphene on Re(0001)}
\subsection*{Dirac cone of quasi free-standing graphene}
\begin{figure*}[!hbt]
	\centering
	\includegraphics[width=16.35cm]{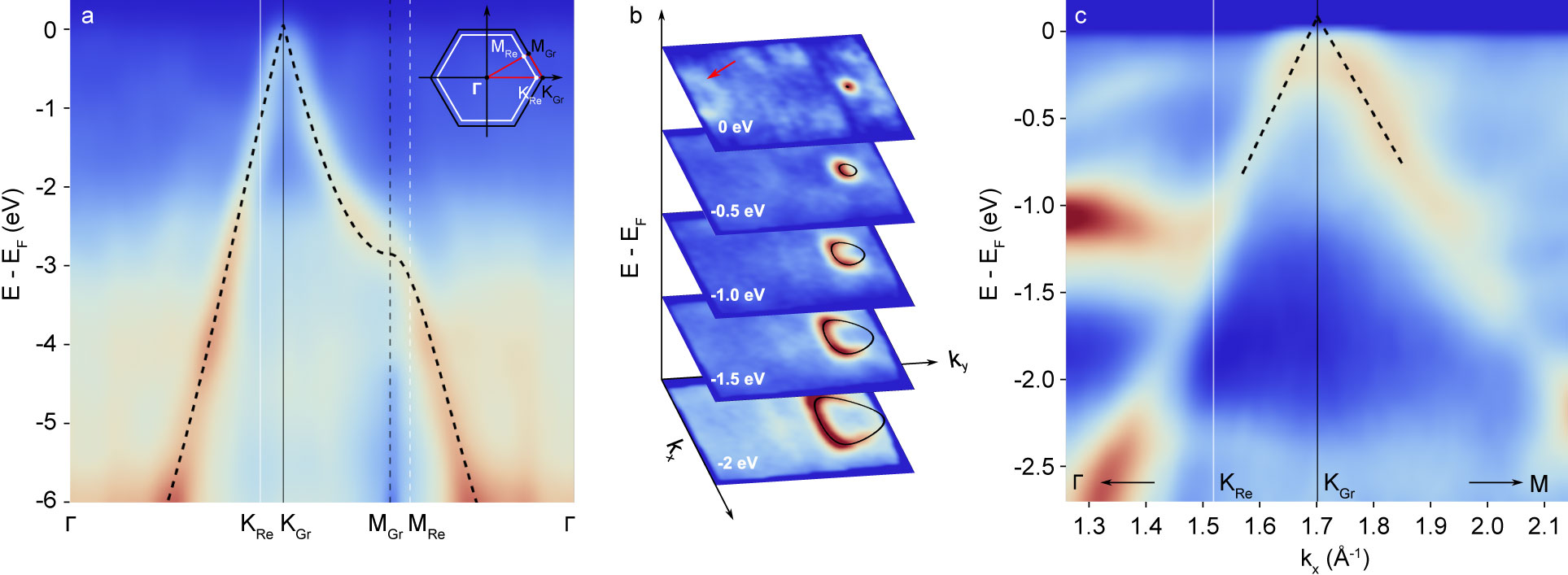}
	\caption{(a) Photoemission intensity plot along the $\Gamma$-K-M-$\Gamma$ direction of Gr/(1.2~ML Au)/Re(0001). Inset: Reciprocal space sketch showing the $\Gamma$-K-M-$\Gamma$ direction (in red). (b) Energy slices of the photoemission intensity, revealing the trigonal warping of graphene $\pi$ state. The presence of a parabolic state at $\Gamma$ with a minimum at about -0.2~eV is indicated with a red arrow. A first-nearest neighbour tight-binding fit of graphene's $\pi$ state is shown on both (a) and (b) plots (black solid lines) \cite{reich2002}. (a) and (b) plots were acquired with an $\mathrm{He}_\mathrm{II}$ source (40.8~eV photon energy). (c) Photoemission intensity plot (zoom around the K points) along the $\Gamma$-K-M direction (neither $\Gamma$ nor M are reached), with an $\mathrm{He}_\mathrm{I}$ source (21.2~eV photon energy). A linear fit of graphene's $\pi$ state is shown (black dashed lines). The Dirac point is found at the intersection of the black dashed lines.}
	\label{fig4}
\end{figure*}

Photoemission measurements were performed on graphene/Re(0001) intercalated with 1.2~ML of Au, to maximize the surface fraction of intercalated graphene. No moir\'e satellite spots are visible in LEED (figure~S5) after intercalation, suggesting close-to-total intercalation of graphene/Re(0001) with Au. Figure~\ref{fig4}(a) displays the electronic band structure along the $\Gamma$-K-M-$\Gamma$ direction of graphene on Re after the intercalation of Au, with a $\mathrm{He}_\mathrm{II}$ source. The $\pi$ band shows a clear linear dispersion, with a crossing (Dirac) point above the Fermi level. No $\pi^*$ band is observed. These observations are reminiscent of others made with related systems \cite{papagno2013,marchenko2012,shikin2013,marchenko2016,gierz2010,fei2015}, and suggest that graphene here has a free-standing character. In contrast, ARPES measurements on graphene/Re(0001) in the absence of the Au intercalant suggest strong electron donation from the substrate to graphene and hybridization between the C and Re orbitals (figure~S1(d)), consistent with previous reports \cite{papagno2013,estyunin2017}.

A first-nearest neighbour tight-binding fit was performed on the $\pi$ state of graphene/Au/Re(0001) (see figure~\ref{fig4}(a)), following the model described in Ref.~\cite{reich2002}. The refined fit parameters are: $\epsilon_{2p}$ = 65~meV, $\gamma_0$ = -2.98~eV and $s_0$ = 0.028, where $\epsilon_{2p}$, $\gamma_0$ and $s_0$ are the on-site energy parameter, the tight-binding hopping parameter and the overlap parameter, respectively.  The fitting procedure shows that the Dirac point is above the Fermi level, \textit{i.e.} graphene is slightly $p$-doped. This finding is consistent with experimental observations on related systems, namely graphene-on-metal systems intercalated with Au, graphene with adsorbed Au adatoms, thin Au films on graphene, and graphene/Au(111), made with ARPES \cite{varykhalov2008,gierz2008,gierz2010,enderlein2010,marchenko2012}, quasiparticle interferences \cite{leicht2014_acsnano,leicht2014_prb}, STS \cite{klusek2009} and Raman spectroscopy \cite{wang2011} measurements, as well as theoretical calculations \cite{giovannetti2008,kang2010,slawinska2011}. In contrast, on graphene/Re(0001), Ag \cite{papagno2013} and Pb \cite{estyunin2017} intercalation resulted in $n$-doping of graphene. 

Figure~\ref{fig4}(c) shows a close-up view of the electronic band structure around the K point, with a $\mathrm{He}_\mathrm{I}$ source. A linear fit was performed on the branches of graphene $\pi$ state (black dashed lines). The Dirac point is found at the intersection of the black dashed lines, 66~$\pm$~40~meV above the Fermi level, in good agreement with the tight-binding fit mentioned before. We observe no sign of moir\'e replica bands or minigaps along the $\pi$ band that would be related to the moir\'e superpotential, contrary to observations in other graphene moir\'e systems \cite{pletikosic2009,sanchez-barriga2012,kralj2011,usachov2012,wang2016,ohta2012}. This once more shows that the interaction between graphene and Au is very weak. Whether a graphene-substrate interaction-related kink is present along the $\pi$ band, that would arise due to the Au layer \cite{varykhalov2008}, cannot be established. The trigonal warping of the $\pi$ band is clearly observed starting from 1~eV below the Fermi level in the $(k_\mathrm{x},k_\mathrm{y})$ cuts of the band structure shown in figure~\ref{fig4}(b).

Besides the Dirac cone, the band structure of gold-intercalated graphene on Re(0001) presents a parabolic state at $\Gamma$ with a minimum at about -0.2~eV. This state was not observed before Au intercalation. It can be seen in figure~\ref{fig4}(b) on the energy slice taken at the Fermi level, indicated with a red arrow. This state may correspond to an electronic state in Au. The Shockley surface state of Au(111) was shown to survive underneath graphene in gold-intercalated graphene on Ir(111) \cite{leicht2014_acsnano}, shifted by $\simeq$~100~meV to lower binding energy compared to the value of -505~meV observed for Au(111) single-crystals \cite{kliewer2000}. This shift was attributed to the finite intercalated Au film thickness ($\simeq$~20~ML of Au). The presence of graphene may also have contributed to the energy shift, as it was demonstrated for the Cu(111) Shockley state \cite{hollen2015}. Here, the electronic state measured in ARPES is shifted by $\simeq$~300~meV with respect to the Shockley state of Au(111) single-crystals, which can be attributed to the presence of graphene and, more importantly, to the reduced thickness of the intercalated Au film (1~ML) compared with Ref.~\cite{leicht2014_acsnano}. We propose that the state observed in ARPES corresponds to the maximum appearing at about -0.08~eV in the density of states of Au (figure~\ref{fig3}(b)). We previously gave an explanation for the absence of the corresponding signature in the experimental local conductance measured with STS.

\subsection*{Raman modes of quasi free-standing graphene}
\begin{figure*}[!hbt]
	\centering
	\includegraphics[width=15.2cm]{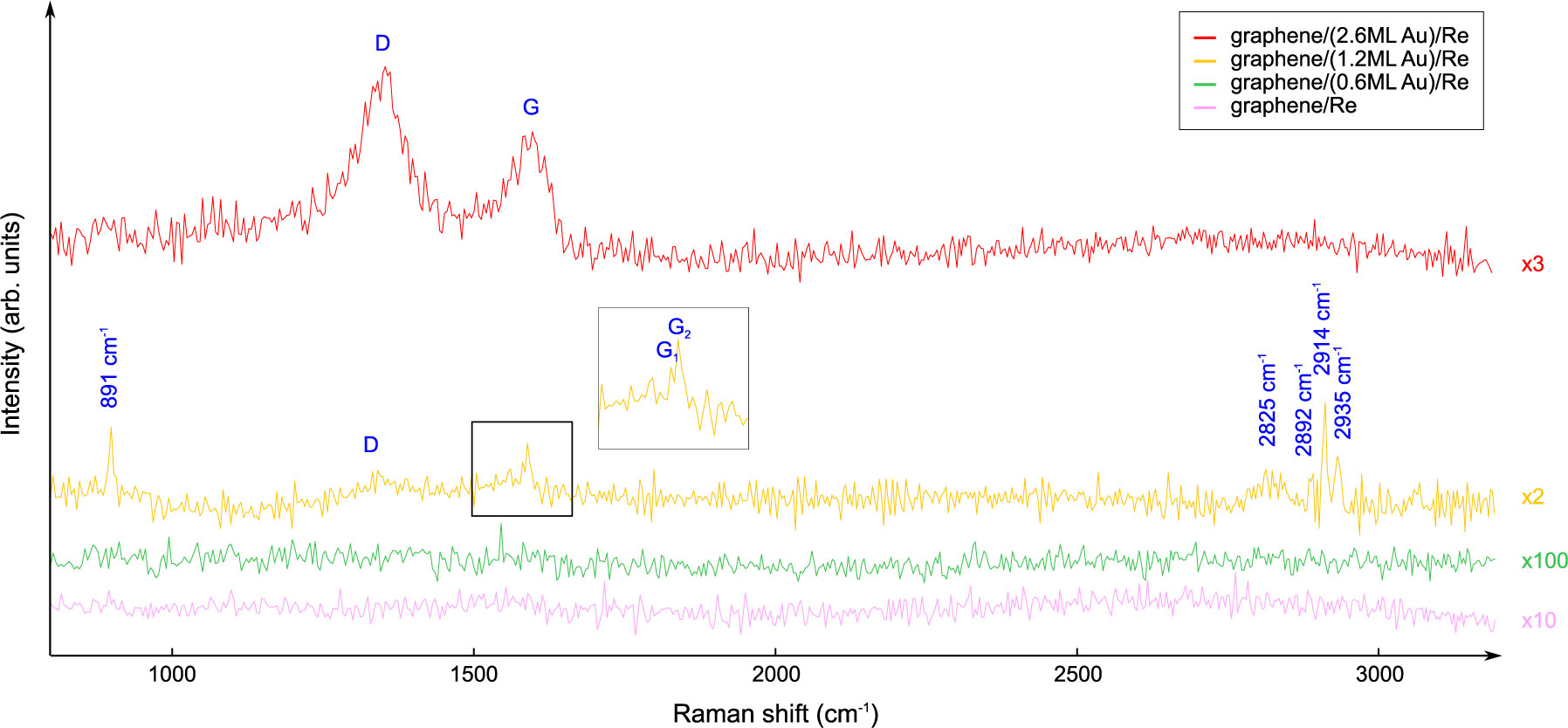}
	\caption{Background-subtracted Raman spectra (532~nm, 1.2~mW/$\mu$m$^2$) of graphene on Re(0001) before and after the intercalation of 0.6~ML, 1.2~ML and 2.6~ML of Au. Raman spectra are offset for clarity. The signal was collected through 50$\times$ and 100$\times$ objectives.}
	\label{fig5}
\end{figure*} 

The observation of quasi free-standing electronic properties in graphene/Au/Re(0001) suggests that the characteristic G and 2D vibration bands of graphene could be observable in Raman spectroscopy. Indeed, unlike in the case of graphene/Re(0001) \cite{papagno2013} or graphene/Ru(0001) \cite{marchini2007,sutter2008,sutter2009-Ru}, the $\pi\leftrightarrow\pi^*$ electronic transitions involved in the Raman processes should be allowed after Au intercalation. Besides, the expected weak interaction between Au and graphene should not alter the Kohn anomalies of the phonon dispersion of isolated graphene, appearing as kinks in the phonon dispersion of the LO mode at $\Gamma$ and of the TO mode at K, which are involved in the Raman G and 2D modes of graphene, respectively \cite{allard2010}. 

Figure~\ref{fig5} shows Raman spectra of graphene on Re(0001) before and after the intercalation of 0.6~ML, 1.2~ML and 2.6~ML of Au. In the explored wavenumber range (800-3200~cm$^{-1}$), the background related to luminescence from the substrate is rather high. A quadratic background was subtracted from all four spectra, to recover comparable baselines. 

We start with a short discussion on the Raman spectrum measured on graphene/Re(0001) (without intercalant). For this particular measurement, graphene growth was performed on a rhenium thin film. As mentioned before, this procedure yields single-layer graphene covering the whole sample surface. No specific Raman signal is observed, and in particular no G or 2D band is detected (figure~\ref{fig5}). This is consistent with the large electronic bandgap found in the ARPES data (figure~S1(d)) between the $\pi$ and $\pi^*$ electronic bands, due to which no electronic transition with sufficiently low energy is available in graphene to make the Raman processes active \cite{ferrari2013}.

The G and 2D bands are not observed either after the intercalation of 0.6~ML of Au, at odds with our expectations for graphene decoupled from its substrate (\textit{via} the Au intercalant). We conclude that the background level is too high and the graphene vibration modes are too weak for this fraction of intercalated graphene.

For larger amounts of intercalated Au (1.2 and 2.6~ML), we do observe some of the typical graphene vibration bands. AFM measurements show that for a nominal 2.6~ML Au deposit, 5-6 ML of Au are locally intercalated (see SI), \textit{versus} 1-2~ML for a nominal 0.6~ML Au deposit (as previously deduced from STM measurements). The G peak appears at 1595~cm$^{-1}$ in (2.6~ML Au)-intercalated graphene on Re. It is split into two peaks in (1.2~ML Au)-intercalated graphene on Re, referred to as $G_\mathrm{1}$ and $G_\mathrm{2}$ in figure~\ref{fig5}, pointing to a lifting of the degeneracy of the E$_{2g}$ mode due to uniaxial strain \cite{huang2009,mohiuddin2009}. The splitting is reproducible: it is observed in many different regions of the sample. The two bands $G_\mathrm{1}$ and $G_\mathrm{2}$ are positioned at 1580~cm$^{-1}$ and 1589~cm$^{-1}$ respectively. The measured positions of the G peaks in (1.2~ML Au)- and (2.6~ML Au)-intercalated graphene on Re are compatible with the slight $p$-doping measured in ARPES, as we expect little to no upshift of the G peak \cite{yan2007,das2008} with respect to undoped graphene (1585-1590~cm$^{-1}$). Besides, the position of the G peak might be slightly overestimated in (2.6~ML Au)-intercalated graphene on Re, since, as it will be discussed later, it is likely that the broad feature appearing at 1595~cm$^{-1}$ is composed of both the G and D' peaks.

Our measurements reveal no 2D band, the overtone of the D band that is active even without defects. Indeed, from the D band wavenumber (around 1350~cm$^{-1}$, see later) the 2D band's wavenumber would be 2700~cm$^{-1}$. We attribute the absence of the 2D band to a broadening and concomitant reduction of intensity of the 2D peak due to defects \cite{lucchese2010,cancado2011} (as we will see later, the level of defects is quite high in intercalated samples). We expect that these effects lead to a 2D band below the background level, hence not detected in practice.

The D peak, which stems from a double-resonance scattering process in the presence of defects, appears at 1349~cm$^{-1}$ and 1351~cm$^{-1}$ for (1.2~ML Au)- and (2.6~ML Au)-intercalated graphene on Re, respectively. These values are consistent with Refs.~\cite{eckmann2013} and \cite{pocsik2008}, where the D peak was measured around 1345~cm$^{-1}$ and 1350~cm$^{-1}$, respectively, at our excitation energy.

We observe other Raman bands, at 891, 2825, 2892, 2914 and 2935~cm$^{-1}$ in graphene/(1.2~ML Au)/Re (figure~\ref{fig5}). It is excluded than any of these five bands corresponds to a first-order or a second-order scattering process involving rhenium phonons \cite{zacherl2010} \footnote{The phonon dispersion curve of rhenium was calculated in Ref.~\cite{zacherl2010}. The energy of the highest energy rhenium phonon is $\simeq$~23~meV, hence a first-order scattering process involving this phonon would result in a 190~cm$^{-1}$ Raman shift.}. These peaks could instead relate to the minority rhenium carbide phase coexisting with graphene. The highest-energy band could also correspond to the D+D' feature in graphene, that is linked to the presence of defects \cite{cancado2011,ferrari2013}.

\section*{Defects in gold-intercalated graphene on Re(0001)}
We now turn to a discussion on the consequences of intercalation on the structural quality of graphene. The width of the Raman bands informs on the quantity of defects in our samples. We choose to focus on (2.6~ML Au)-intercalated graphene on Re, as the signal-to-noise ratio is higher and hence allows a quantitative analysis. Cançado \textit{et al.} introduced two methods for assessing the inter-defect distance $L_\mathrm{D}$ based on the analysis of the D and G peaks \cite{cancado2011}. First, we extrapolate the width of the D band we measured with a 532~nm laser excitation (87~$\pm$~6~cm$^{-1}$) to the range explored by Cançado \textit{et al.}, assuming a linear dependence with wavenumber \cite{eckmann2013}. The G band width (70~$\pm$~7~cm$^{-1}$) being non-dispersive, does not need to be extrapolated. These widths are higher than the highest values reported in Ref.~\cite{cancado2011} for graphene subjected to Ar$^+$ ion bombardment, suggesting that $L_\mathrm{D}$ is below 2~nm.

To obtain a better estimate of $L_\mathrm{D}$ we turn to the second method described by Cançado \textit{et al.}, which was also proposed by Lucchese \textit{et al.} \cite{lucchese2010}. The relative intensity of the G and D bands $I_\mathrm{D}/I_\mathrm{G}$ can be used to assess the fraction area around defects where the scattering processes responsible for the D mode occur \cite{lucchese2010,cancado2011}. In turn, this provides an estimate of $L_\mathrm{D}$. Here again we must extrapolate the $I_\mathrm{D}/I_\mathrm{G}$ value of 1.4~$\pm$~0.1 that we measured with a 532~nm source to the range covered in Ref.~\cite{cancado2011}. Using the wavelength-$I_\mathrm{D}/I_\mathrm{G}$ linear dependence determined in Refs.~\cite{eckmann2013,barros2007}, we deduce $L_\mathrm{D}\simeq$~1~nm. 

Finally, using the quadratic form proposed by Ferrari \textit{et al.} \cite{ferrari2001,ferrari2013}, we derive $L_\mathrm{D}\simeq$~1.5$~\pm$~0.1~nm.

All three estimates of the distance between defects seem consistent and correspond to a $10^{14}$~cm$^{-2}$ defect density. Obviously, these estimates apply only to Raman-active defects. For such a high level of defects, one can consider that the broad feature appearing at 1595~cm$^{-1}$ is actually composed of both the G and D' peaks \cite{lucchese2010,cancado2011}. The D' peak is a disorder-induced peak located around 1620~cm$^{-1}$. Interestingly, our estimates of $L_\mathrm{D}$ are close to the ones we deduce, using the same three methods, in related systems, namely graphene/Ru(0001) intercalated with Pb \cite{fei2015} and graphene/SiC(0001) intercalated with Au \cite{gierz2010}.

What the source and nature of the defects are is a natural question. Graphene on Re(0001) already contains a certain amount of defects prior to intercalation \cite{artaud2018}. These defects are first the boundaries that surround single-crystal graphene grains with size of the order of few tens of nanometers. Other kinds of defects may be point defects, \textit{e.g.} C vacancies. The intercalation process itself may induce defects such as monovacancies or divacancies in graphene \cite{sicot2014}, \textit{via} for instance a strong interaction between Au monomers and graphene \cite{boukhvalov2009}. A Cu penetration path beneath graphene was proposed to occur \textit{via} metal-aided defect formation with no or poor self-healing of the graphene sheet for the Cu intercalation of graphene/Ir(111) \cite{sicot2014}. Point defects in graphene \cite{coraux2012,albalushi2016,sicot2012,petrovic2013,vlaic2014} and graphene edges \cite{vlaic2014,sutter2010,vlaic2017,vlaic2018} have also been proposed to be potential intercalation pathways. Another mechanism, based on the material diffusion \textit{via} metal-generated defects, followed by the subsequent healing of the graphene lattice, has been suggested \cite{huang2011,sicot2012}.

Figure~\ref{fig2}(c) shows an atomically-resolved STM image taken on a Au-intercalated graphene region, revealing a damaged graphene sheet at the atomic scale. Carbon monovacancies are indicated with white arrows. We detect no such defects in graphene/Re(0001) (see Ref.~\cite{tonnoir2013} and figure~S1(a)). Other defects similar to the one visible in the center of the STM image are observed in intercalated regions. They may correspond to defects in the intercalated Au layer. Our STM observations support a typical distance of the order of 1~nm between point defects in intercalated graphene. We propose that these defects were created upon intercalation. 

\section*{Superconducting properties}
\begin{figure}[!hbt]
	\centering
	\includegraphics[width=6.6cm]{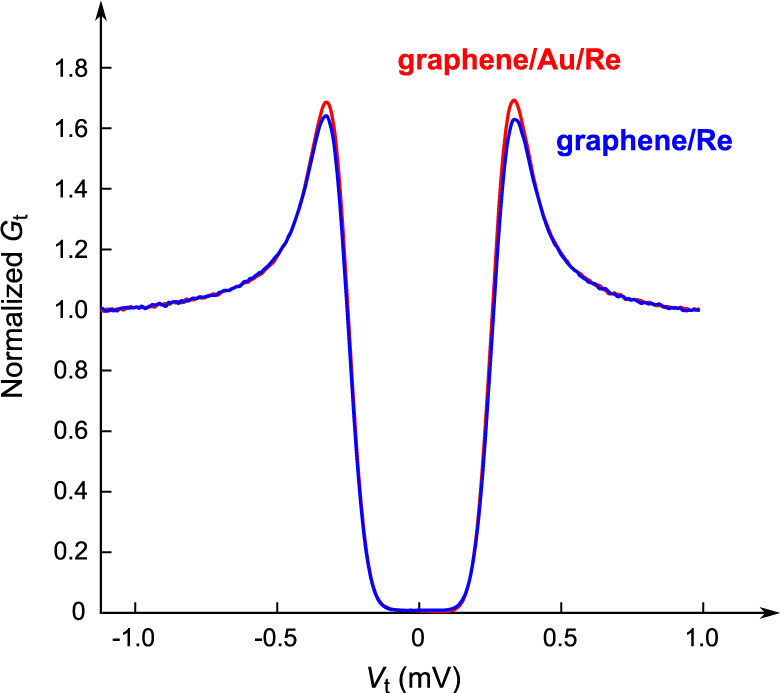}
	\caption{Normalized low-energy conductance $G_\mathrm{t}(V_\mathrm{t})$ spectra measured on graphene/Re and graphene/Au/Re. T = 65~mK.}
	\label{fig6}
\end{figure}

Figure~\ref{fig6} shows normalized conductance spectra measured at low bias voltage and at 65~mK on graphene-covered Re(0001) and gold-intercalated graphene on Re(0001). The conductance spectra are hardly discernible from one another and can be well fitted with the standard Bardeen-Cooper-Schrieffer (BCS) model. The same superconducting gap $\Delta$ = 280~$\pm$~10~$\mu$eV is observed uniformly on the surface, on both intercalated and non-intercalated regions. This suggests that the interface resistance in both regions is sufficiently low so that the proximity-induced superconducting gap is upper bounded, in both regions, by its value for bare Re. Despite the fact that our STS data cannot tell us more on the quality of the interfaces after gold intercalation, it is consistent with the counterintuitive result that for extended surface contacts like our fully covered surface, the interface transparency is higher for metals weakly bounded to the graphene \cite{nemec2008}. In any case, intercalation of gold in between graphene and Re(0001) does not alter the rhenium-induced superconductivity in graphene.

The superconducting gap measured here is slightly different from the one reported in Ref.~\cite{tonnoir2013} (330~$\pm$~10~$\mu$eV) on a graphene-covered Re(0001) thin film. The value reported in Ref.~\cite{tonnoir2013} is rather high compared with the superconducting gap measured on bare Re thin films ($\simeq$~255~$\mu$eV \cite{dubouchet2010}). It was attributed to the presence of C atoms dissolved into the bulk of the Re thin film, as a result of the bulk saturation/segregation process at the origin of graphene growth. For graphene grown on Re(0001) single-crystal (our sample), the substrate can be viewed as an infinite reservoir of rhenium atoms, hence we do not expect to saturate the Re bulk with dissolved C atoms. The superconducting gap we measure is therefore closer to the value measured for bare Re. 

\section*{Summary and concluding remarks}
We have investigated the structural, electronic and vibrational properties of graphene on Re(0001) intercalated with Au, by means of RHEED, LEED, AFM, STM/STS, ARPES, Raman spectroscopy measurements, and DFT calculations. Our data consistently show that graphene recovers a free-standing character after Au intercalation, in sharp contrast with the strongly hybridized graphene-on-Re(0001). This is apparent in the electronic band structure, in the flattening of the graphene topography, in the disappearance of the moir\'e-related streaks in diffraction measurements, and in the vibrational spectra of graphene. The weak corrugation observed in STM, as well as the absence of minigaps and replicas in ARPES indicate a weak graphene-Au interaction. The vibrational spectra allowed us to assess the density of defects in the systems, which is high, with a typical distance between phonon scattering centers of the order of 1~nm. This value, and the ones we extracted from previous data published for other systems, indicate that a high density of defects is created in graphene during the process of intercalation. Obviously, this is a serious issue often overlooked in works relying on intercalated graphene, which calls for further, specific efforts devoted to the optimization of the intercalation process. We finally prove superconductivity in graphene, induced by a remote proximity effect through the Au layer, from the Re(0001) substrate. Graphene/Au/Re(0001) appears as an ultra-clean platform, ideally suited to address, for example, the competition between superconductivity and magnetic orders. In this context, this new platform could allow testing the predictions of Yu-Shiba-Rusinov states \cite{shiba1968,wehling2008,lado2016}, or manipulating such states \cite{sanjose2015}.

\section*{Acknowledgments}
This work was supported by the R\'egion Rhône Alpes (ARC6 program) and the Labex LANEF. Funding from the French National Research Agency under the 2DTransformers (OH-RISQUE program ANR-14-OHRI-0004) and ORGANI’SO (ANR-15-CE09-0017) projects is gratefully acknowledged. We thank Amina Kimouche for sample preparation.

\section*{References}

\end{document}